\documentclass{aa}

\usepackage[dvips]{graphicx}
\usepackage{verbatim, subfigure}
\usepackage{apalike} 
\usepackage{amsmath} 
\usepackage{dcolumn}
\usepackage{longtable}

\newcommand{\vald}{{\sc vald}}                              
\newcommand{\vsini}{\ensuremath{v\sin i}}                   

\newcommand{\kms}{\ensuremath{\text{km\,s}^{-1}}}           

 \newcommand{\vwa}{{\sc vwa}}         
 \newcommand{\iraf}{{\sc iraf}}         

\def\iet{\,{\sc i}} \def\ii{\,{\sc ii}}  

\newlength{\dplotwidth}
\begin{document}


 \title{Abundance analysis of targets for 
          the COROT / MONS asteroseimology missions}

   \subtitle{I. Semi-automatic abundance analysis of the $\gamma$ Dor star HD~49434\thanks{Based on
           observations obtained at Observatoire d'Haute Provence, France
                 and at the Observatory of Sierra Nevada, Granada, Spain}}

   \author{H.~Bruntt\inst{1}
        \and
         C.~Catala\inst{2,5} 
        \and
         R.~Garrido\inst{3}
        \and
         E.~Rodr\'\i{}guez\inst{3}
        \and
          J.~C.~Bouret \inst{4} 
        \and
          T.~Hua \inst{4} 
        \and
          F.~Ligni\`eres \inst{2} 
        \and
          S.~Charpinet \inst{2} 
        \and
          C.~Van't Veer-Menneret \inst{5} 
        \and
          D.~Ballereau \inst{5}}

   \offprints{H. Bruntt, \email{bruntt@ifa.au.dk}}

   \institute{Institute for Physics and Astronomy, University of Aarhus,
              Bygn.~520, DK-8000 Aarhus C, Denmark
        \and 
             Laboratoire d'Astrophysique de l'OMP, CNRS UMR 5572,
             Observatoire Midi-Pyr\'en\'ees, 14, avenue Edouard Belin, F-31400
             Toulouse, France 
        \and 
              Instituto de Astrof\'\i{}sica de Andaluci\'a, Espan\~a
        \and
             Laboratoire d'Astrophysique de Marseille, France
        \and
             Observatoire de Paris, France 
}


\date{Received ..... / Accepted .....}

\abstract{One of the goals of the ground-based 
support program for the COROT and MONS/R\o mer satellite missions
is to select and characterise suitable target stars for the
part of the missions dedicated to asteroseismology.
While the global atmospheric parameters may be determined 
with good accuracy from the Str\"omgren indices, careful
abundance analysis must be made for the proposed main targets.
This is a time consuming process considering 
the long list of primary and secondary targets.
We have therefore developed new software called \vwa\ for this task.
The \vwa\ automatically selects
the least blended lines from the atomic line database \vald, and 
consequently adjusts the abundance in order to find the best match 
between the calculated and observed spectra. 
The variability of HD~49434 was discovered as part of COROT
ground-based support observations. Here 
we present a detailed abundance analysis of HD~49434 using \vwa.
For most elements we find abundances somewhat below
the Solar values, in particular we find [Fe/H] $= -0.13 \pm 0.14$.
We also present the results from the study of the variability 
that is seen in spectroscopic and photometric time series 
observations. 
From the characteristics of the variation seen in photometry and
in the line profiles we propose that HD~49434 is a variable star 
of the $\gamma$ Doradus type.
   }

\maketitle


\section{Introduction}

We first observed HD~49434 (F2V-type) with the 
ELODIE spectrograph in December 1998 as part of a program with the 
aim of characterising and selecting potential targets for the 
asteroseismology space mission COROT (Catala 2001). 
The strategy is to obtain a number of high resolution, 
high signal-to-noise spectra of each potential target. 
For any stars that present line asymmetries or
variable line profiles we then carry out more complete monitoring.

Before these observations the star was not considered to be variable,
but we discovered systematic low-amplitude variations of the line 
profiles. Since then we have carried out more extensive 
spectroscopic and photometric observations.

The purpose of the present paper is to derive the global
atmospheric parameters of HD~49434 and to perform an analysis of
its photometric and spectroscopic variations. 

The French COROT as well as the Danish MONS/R\o mer satellite 
are both asteroseismology missions which will study mostly 
solar-like stars. 
Before the launch of these missions (around 2004--5) 
we need an accurate estimation of the atmospheric parameters and 
abundances of the proposed target stars. 
First of all stars with certain parameters may not be suitable for 
a detailed asteroseismological study, eg.~chemically peculiar stars or 
stars with high rotation rate. Also, in order to carry out the
detailed asteroseismological modelling of the stars 
we need to be able to constrain as many of the global parameters 
of the stars as possible.
Around 10 and 25 primary targets will be chosen for 
COROT and MONS/R\o mer, respectively. It is a very
time-consuming process to carry out the abundance analy\-sis 
of all potential targets manually and we have therefore developed a fast 
semi-automatic abundance analysis procedure called \vwa, which 
will be described here. The result for HD~49434 
is presented, and we will also make comparisons with stars for 
which abundance analyses have already been carried out manually.

In Section~\ref{specobs} we describe the spectroscopic observations 
and in Section~\ref{abundance} we
describe the \vwa\ software and the result of the abundance analysis.
In Section~\ref{photometry} we describe the analysis of the
photometric time series, while in Section~\ref{lsd} 
we analyse the line profile variations in the spectroscopic time series.


\section{Spectroscopic observations \label{specobs} }

HD~49434 is an F2V-type star of magnitude $V = 5.74$ (cf.~Table~\ref{stromgren_tab}).
We monitored the star spectroscopically from Observatoire de Haute-Provence,
using the ELODIE spectrograph, attached to the 1.93\,m telescope. The
spectrograph is a cross-dispersed spectrograph, providing a complete 
spectral coverage of the 3\,800--6\,800~\AA~region, 
with a resolving power of about $R = 45\,000$ (Baranne et al.\ 1996). 

The first spectroscopic observations of HD~49434 were carried out 
in December 1998 as part of the search for potential COROT targets. 
The spectra revealed the presence of variable asymmetries of low amplitude 
in the line profiles, indicative of non-radial pulsations. 
Hence, the star was monitored more thoroughly in subsequent observing runs.

On January 26, 2000, HD~49434 was monitored continuously for 90 minutes, with 
one 5 min spectrum every 7 minutes. The weather and seeing 
conditions were very good, and the average S/N ratio at 6\,000~\AA~is
250 per velocity bin of 10 \kms.

The star was re-observed on December 10, 2000, almost continuously for 
320 minutes, during which as many as 51 five-minute spectra were recorded. 
The observing conditions of this second monitoring were not as good as the 
previous one, so that the resulting average S/N ratio at 6\,000~\AA~is only 
about 170 per pixel per velocity bin of 10 \kms.

The data used for the abundance analysis of HD~49434 are the spectra 
obtained during the best weather conditions: 12 spectra from 
January 26, 2000. On the other hand,
all of the observed spectra were used for the study of the variability 
of line profiles. 
Since the data analysis of the spectra was made by two groups independently, 
we will describe the data reduction in separate sections, 
i.e.~abundance analysis in Section~\ref{abundance-datareduction} 
below and the variability of line profiles 
in Section~\ref{lsd-datareduction}.


\begin{figure*}\centering
  \begin{center}
    \caption{Observed (thin line) and synthetic (thick line) spectrum 
around twelve lines which were used for the abundance determination.
The lower portion of the each plotted line shows the relative 
difference between the observed and synthetic spectrum. The name of the 
element and the central wavelength (in \AA ngstr\o m) of each fitted line 
is given at the top of each plot\label{lines_fig}}  
\includegraphics[height=3.7cm,width=\textwidth]{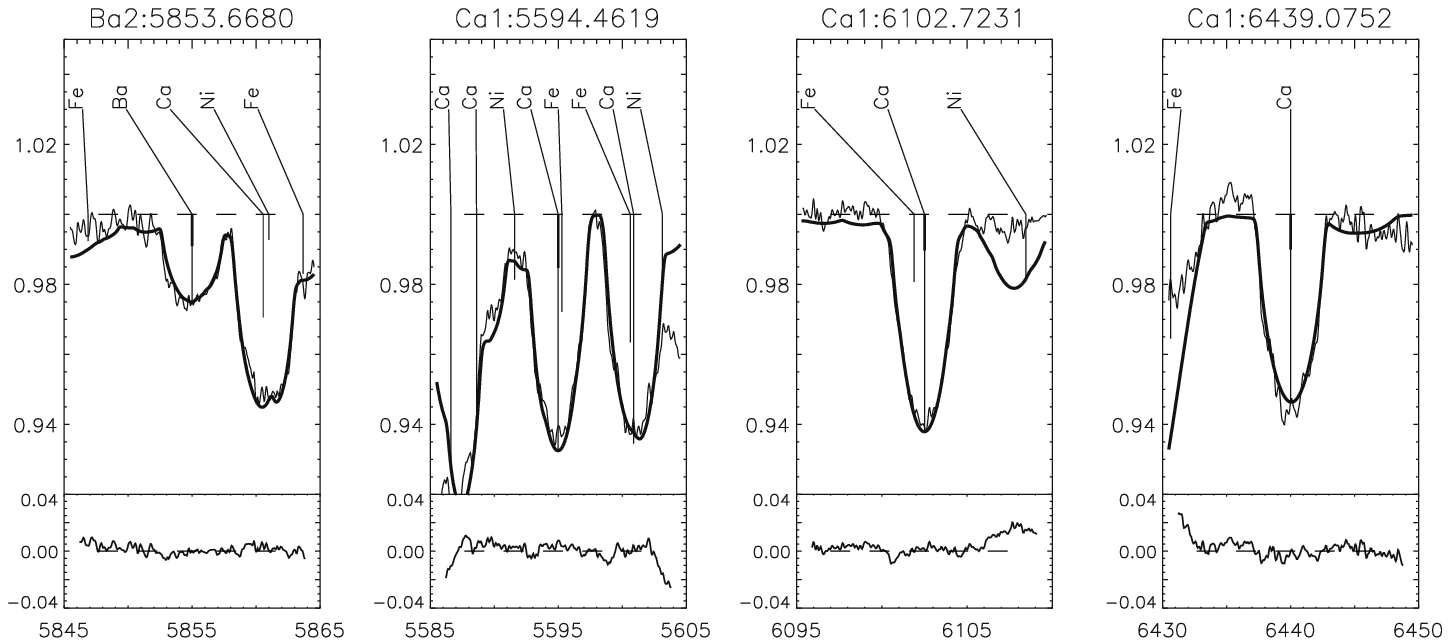}
\vskip 0.55cm
\includegraphics[height=3.7cm,width=\textwidth]{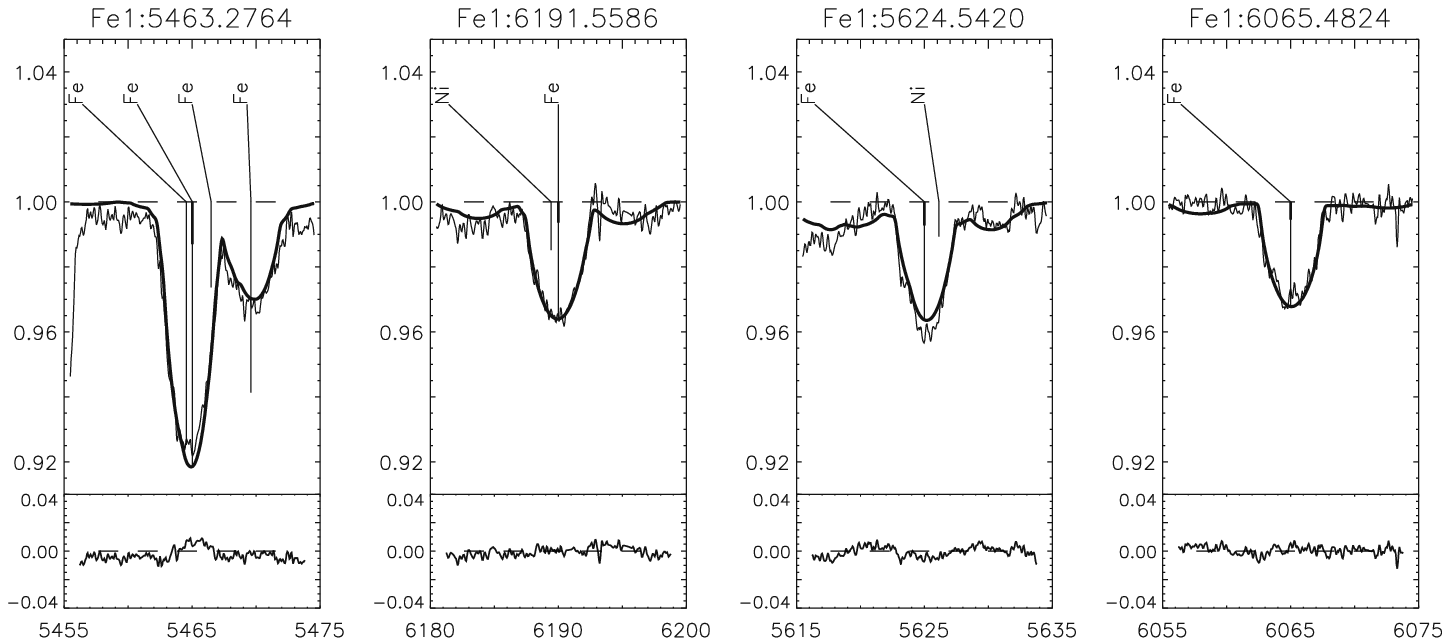}
\vskip 0.55cm
\includegraphics[height=3.7cm,width=\textwidth]{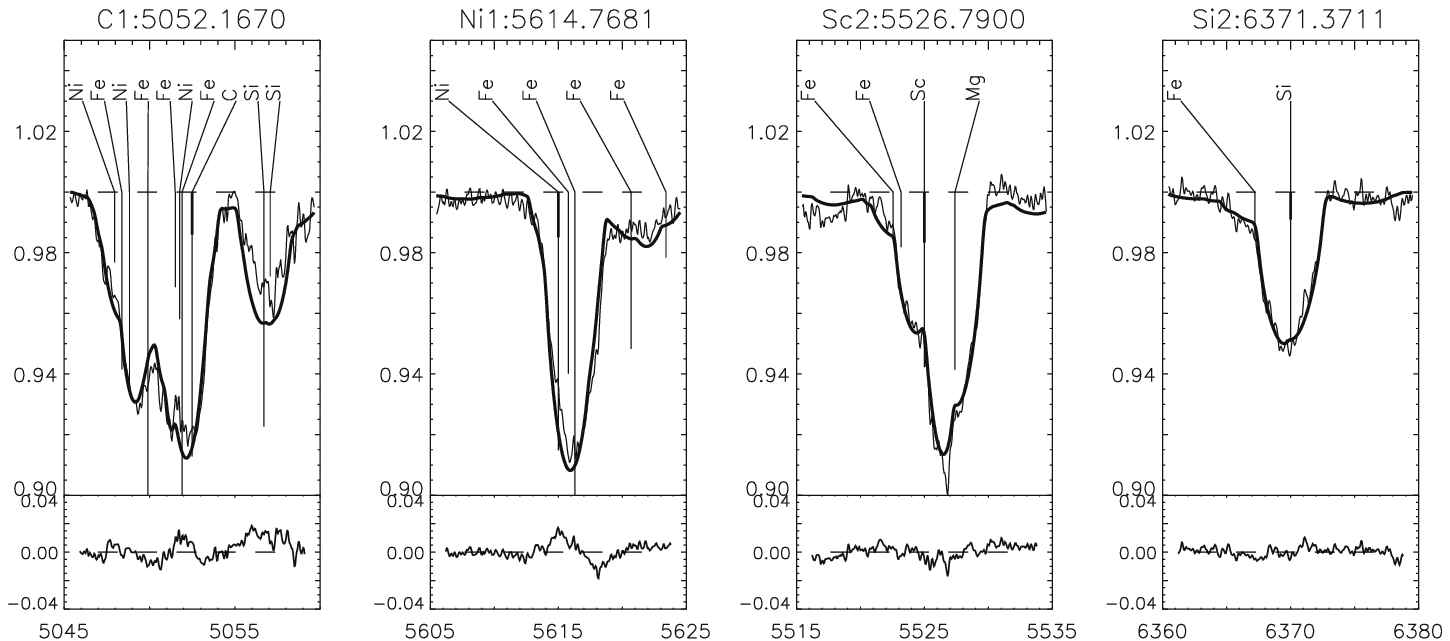}

  \end{center}
\end{figure*}


\section{Abundance analysis \label{abundance}}

In this section we will describe the abundance analysis of
HD~49434. For this purpose we have developed new software for the
semi-automatic analysis of spectra. 
The abbreviation for the software is \vwa\ which explains
what the program can determine from the spectrum, 
i.e.\ $v \sin i$ of the star, 
wavelength shift (due to radial velocity), and abundance analysis. 

\subsection{Data reduction\label{abundance-datareduction}}

The bias subtraction, flat fielding, scattered light subtraction,
and extraction of the spectral orders were done with \iraf. 
The wavelength calibration was made 
using a combined Th-Ar comparison spectrum. We made low-order cubic 
spline fits to make the best continuum estimate. 
We made sure that the overlap between orders was better than 0.5\%. 
We note that we have not used the region near the hydrogen lines for 
abundance analysis. The hydrogen lines cover several spectral orders and
thus a determination of the continuum level is difficult.
One possibility to overcome this problem is to use the average of the 
continuum level for orders adjacent to the hydrogen lines. 
This has been done by Lastennet et al.\ \cite{lastennet} 
and they have estimated the effective temperature from the
H$\alpha$ line to be $T_{\rm eff} = 7250\pm250$~K.

The signal to noise of the combined spectrum (12 spectra) at the
maximum of a spectral order around 6\,000 \AA~is S/N = 400 per 
velocity bin of 10 \kms.

\subsection{Automatic abundance analysis with \vwa}

The \vwa\ package can perform three different tasks: Automatic
selection of lines for the determination of $v \sin i$, determination
of any systematic shift of lines (due to radial velocity), and 
abundance analysis -- the last being the most extensive 
part of the program.

The first two tasks are not fully implemented in \vwa, but $v \sin i$
and $v_{\rm rad}$ can be determined
quite easily by comparison of the observed spectrum
and several synthetic lines: We convolve the
synthetic spectrum with the instrumental profile
and different rotation profiles and determine the best fit
(we use the program {\sc rotate} by Piskunov 1992). 
We find $v \sin i = 84\pm4$ \kms\ which is in agreement 
with Lastennet et al.\ \cite{lastennet} who found 
$v \sin i = 79$ \kms\ from the same data set.
From the wavelength shifts of non-blended lines we
determine the helio-centric radial velocity to be
$v_{\rm rad} = -12.3\pm1.0$ \kms.
Gernier et al.\ (1999) found $v_{\rm rad}=-14.0\pm0.8$ \kms\ from 
Hipparcos data. 
We remove the wavelength shift caused by the radial velocity,
thus the wavelength scale is the same as the laboratory system. 

\subsubsection{Input: model atmosphere and atomic data}

To use \vwa\ one must calculate an appropriate model for 
the atmosphere of the star. We have used a modified version of 
the {\sc atlas9} code (Kurucz~1993) as described by Kupka (1996) -- see
also Smalley \& Kupka (1997).
In this version of {\sc atlas9} the turbulent convection theory
in the formulation of Canuto \& Mazzitelli (1991, 1992) is
implemented.

To calculate the model one needs to know the basic 
parameters of the star, i.e.~$T_{\rm eff}, \log g$, and [Fe/H].
These parameters may be estimated from the Str\"omgren indices of the star
by using the program {\sc templogg} (Rogers~1995) 
which automatically chooses the
most appropriate among several published photometric calibrations.
To determine the basic atmospheric parameters of the star
{\sc templogg} then interpolates the de-reddened photometric indices in
the atmospheric model grid by Kurucz (1979) with the improvements 
suggested by Napiwotzki et al.\ (1993).

For the abundance determination we make use of a complete list of 
atomic line data for the entire observed spectral range. 
The line list is extracted from the \vald\ database (Kupka et al.\ 1999, 
Ryabchikova et al.\ 1999) where all lines deeper than 1\% of the continuum 
are included. From the line list (about 13\,000 lines for HD~49434) 
the least blended lines will be selected by \vwa.

The selection consists of two steps: 1) Selection based on 
the analysis of the degree of blending based on the information in
the line list from the \vald\ database. 
2) These lines are then analysed in the observed spectrum and any
``suspicious'' lines are rejected. 
These steps are described in the following two sections.

\subsubsection{Selecting lines: atomic line list}

For each atomic line extracted from the \vald\ database
we examine the line depths of the neighbouring lines. For a given
line we examine lines within a certain range specified by the user; 
for HD~49434 we have used $3.0 \times$FWHM of an unblended line 
($=3.0 \times 2.2$~\AA $= 6.6$ \AA). The line depths of the neighbouring
lines are convolved by a Gaussian with a FWHM of $1.5 \times$FWHM of an 
unblended line, i.e.~the line depths of lines far away from 
the central line are suppressed.
The selection of lines from the \vald\ line list is based on the 
degree of blending which is described by three parameters. The user
may set his own limits for these parameters, and thus determine
which lines are selected. The parameters are: 1) The depth
ratio of the central line to the deepest (non-convolved) line. This value
must be greater than 1 but can specified by the user (we used 1.35).
2) The number of neighbours with a (convolved) line depth of 25\% of the
central line must be lower than some limit (we used 5). 3) The sum of the
(convolved) line depths must be lower than some number (we used 0.7).




\begin{table*} \centering \footnotesize
  \caption{Abundances of elements in HD~49434. 
           The weighted mean abundance is given as the logarithm of number
           of atoms of a given element to the total number of atoms.
           The weighted standard deviation of the mean is given in parenthesis,
           $n$ is the number of lines that were used, while 
           the last two columns are
           the abundances measured for the Sun 
           (Grevesse \& Sauval 1998) and the difference between 
           the Sun and HD~49434\label{abundances_tab}}
  
  \begin{tabular}{lc|lr|lc}
  \hline 
  Element & Ion & $\log N/N_{\rm tot}$ & $n$ & $({\log N/N_{\rm tot}})_\odot$ & $\Delta$ \\
  \hline


C  & \iet\  & $-3.46(0.13)$ & 1 & $-3.52$ & $+0.06$ \\
Mg & \iet\  & $-4.55(0.07)$ & 3 & $-4.46$ & $-0.09$ \\
Si & \iet\  & $-4.64(0.08)$ & 2 & $-4.49$ & $-0.15$ \\
Si & \ii\ & $-4.57(0.06)$ & 3 & $-4.49$ & $-0.08$ \\
Ca & \iet\  & $-5.51(0.06)$ & 7 & $-5.68$ & $+0.17$ \\
Sc & \ii\ & $-8.89(0.11)$ & 2 & $-8.87$ & $-0.02$ \\
Ti & \ii\ & $-7.16(0.07)$ & 2 & $-7.02$ & $-0.14$ \\ 
Cr & \ii\ & $-6.45(0.05)$ & 4 & $-6.37$ & $-0.08$ \\
Fe & \iet\  & $-4.67(0.02)$ & 16& $-4.54$ & $-0.13$ \\
Fe & \ii\ & $-4.43(0.07)$ & 2 & $-4.54$ & $+0.11$ \\            
Ba & \ii\ & $-9.75(0.09)$ & 3 & $-9.91$ & $+0.16$ \\

\hline

\end{tabular}
\end{table*}

\subsubsection{Rejecting lines: the observed spectrum}

When the lines have been selected from the list of 
atomic lines \vwa\ analyses the {\em observed}
spectrum in the region around each selected line. The aim is to
automatically reject lines that are not suitable for abundance
analysis. This is done by fitting a Voigt profile to the observed spectrum.
A number of ``selection parameters'' specified by the user determines
if a line will be used: the minimum depth of the fitted profile, 
the maximum and minimum width of the profile, 
the maximum value of $\chi^2$ for the profile fit, and
the maximum flux asymmetry between the left and right part of 
the observed line.

Optionally the user is presented with a plot of the observed spectrum
in a region around each of the selected and rejected lines.
The user may then interactively choose if a rejected (accepted) line
should be included (rejected) in the abundance analysis.
In practice this part of the \vwa\ program creates 
two lists of the lines that will be rejected / accepted in the
final selection of lines to be used for abundance analysis, i.e.~these
line lists over-rule the automatic selection.
The line lists may also be edited manually if, for example, 
the user already knows which lines he wants to use for the 
abundance analysis (in accepted line list) -- or
if the user knows that the atomic parameters of certain lines are 
known to be uncertain (in rejected line list).

\subsubsection{Two methods for abundance determination}

The abundance determination is made for each of the selected lines.
It can be done by either of two methods: 1) By the measurement of 
equivalent widths 
of lines in the observed spectrum which are then given
as input to the {\sc width9} code by Kurucz~\cite{atlas9}. 
2) By the adjustment of the abundance of the element of the 
central line of the blend until the computed 
and observed spectrum around each line match, i.e.~they have the
same equivalent width. For the computation of the synthetic 
spectrum we used the program {\sc synth} (version 2.5) by N.~Piskunov 
(see Valenti \& Piskunov~1996).

The first method can only be used for stars where blending is not a
problem, i.e.~slowly rotating stars ($v \sin i < 15$ \kms). 
The second method is much slower but it is the only option for 
fast rotators like HD~49434.

\subsubsection{Deriving the abundances\label{sensitivity}}

For this study we have used the fitting of the abundance in 
the synthetic spectrum in a region 4 \AA~on both sides of the 
central line (the typical FWHM of a single line is about 2.2 \AA~at
6\,000 \AA). After all lines have been fitted the user is presented
with a plot of the region around each line in which both the synthetic
and observed spectrum is shown. Also, all lines with a line depth 
deeper than 25\% of the central line are clearly marked. 
Several fitted lines are shown in Figure~\ref{lines_fig}.
In this ``visual inspection'' part of \vwa\ the user can click 
on different sections of the plot which determines if the user
thinks the line is fitted correctly, 
if there seems to be a problem with the continuum level, 
if the line is severely blended, if the line should simply be ignored,
or if the automatic fit has failed. In the last case these lines can
be fitted interactively by a separate program ({\sc vwa-interact}). 

The abundance of each element is then calculated using only the lines 
accepted by the user. Additional constraints on which lines to use
may be imposed here, i.e.~to use only the least blended lines. For this
the ``sensitivity'' parameter of the line 
can be used to select the best lines or to give higher
weight to certain lines. We define the sensitivity parameter as the 
change in equivalent width when increasing the abundance, 
i.e.~for a non-blended line this would be the slope of the ``curve of growth''
near the abundance of the fitted line. 
The parameter is measured for each line during the automatic 
determination of the abundance and we find that it depends both on 
the properties of the line and the degree of blending.

To be specific, the sensitivity, $S_i$, for each line $i$, is defined as
$S_i = \Delta ({\rm EQW}) / \Delta (\log N / N_{\rm tot})$.
Typical values lie in the range 0.3--1.2 m\AA / dex. 
For the present data we assume that the measured equivalent 
widths of all lines are measured with an accuracy of 5\%, hence the 
error estimate of the abundance for a given line lies in the 
range 0.17--0.04 dex. To get a realistic error estimate for each line one 
must include systematic errors due to eg.~erroneous $\log gf$ values
(i.e.\ oscillator strengths), 
wrong determination of the continuum level, 
the possibility of erroneous abundances (or $\log gf$ values) of 
the lines that blend with the line, and uncertain stellar parameters 
used for the computation of the atmosphere model.

When calculating the abundance of each element based on 
several lines we assign weights to each line which depend on the 
sensitivity of the line. When computing the weights we  
have added an additional error of 0.07 dex which is
the contribution from systematic errors as was discussed above.
For the final abundance result of each ele\-ment we calculate 
the weighted mean, where the weights are given by 
$W_i^{-1} = (0.05 / S_i)^2 + (0.07)^2$. The standard error on the 
derived abundance is also calculated using these weights.

\subsubsection{Reliability of \vwa}

We have tested the program on two stars for which careful
analyses have already been made: The Sun 
(Grevesse \& Sauval 1998) and FG Vir 
(A-type star, $v \sin i = 21.3\pm1.0$ \kms; see Mittermayer~2001). 
For these stars we find good agreement with the previously 
derived abundances, i.e.~within 0.1 dex for all elements. 

We note that two future papers which make use of
\vwa\ are in preparation, i.e.~Rasmussen et al.\ \cite{ngc6134spec} and 
Bikmaev et al.\ \cite{hd32115}. In the later paper 
we study two A-type stars and both have low $v \sin i$ (9 and 21 \kms).
Thus we may compare the result of \vwa\ with a ``classical'' 
abundance determination,
i.e.\ by measuring equivalent widths of non-blended lines. 
The two methods agree within 0.03 dex for all elements. 
We conclude that \vwa\ is indeed a reliable tool for fast
semi-automatic abundance analysis.


\begin{figure*}\centering
  \begin{center}
    \caption{
Differential magnitudes of HD~49434 minus comparison star HD~48922 in the 
Str\"omgren $b$ band (asterisks). 
The small points are magnitude differences between the two comparison
stars HD~49933 and HD~48922 (shifted to allow comparison). Individual
days are marked in every panel as the corresponding HJD\label{strom1_fig}}  
 \includegraphics[height=14.8cm,width=8cm,angle=270]{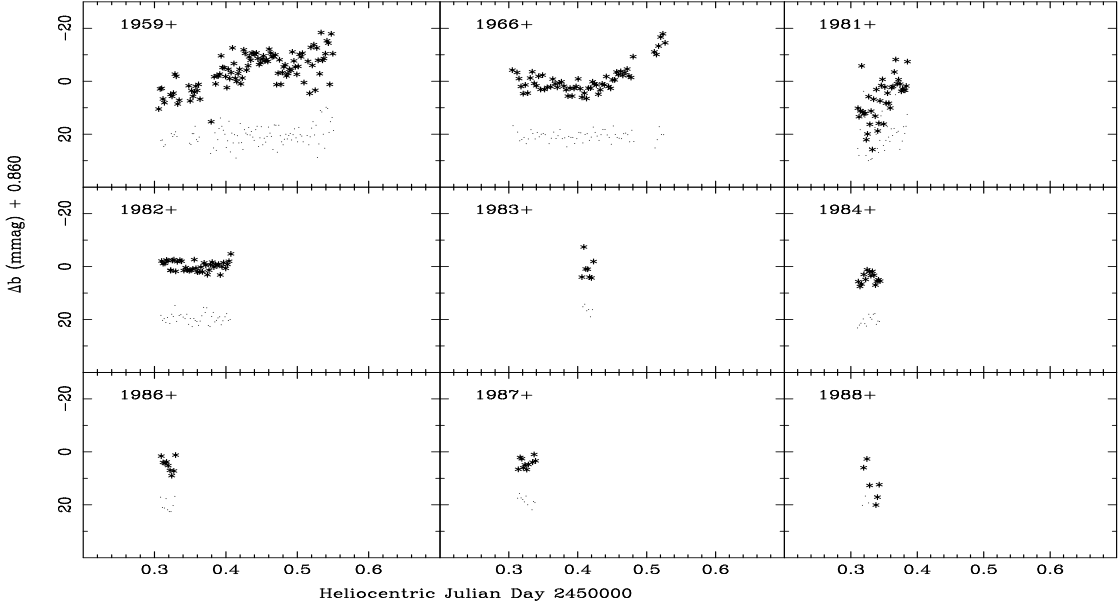}
  \end{center}
\end{figure*}


\subsection{Abundances for HD~49434}

We have estimated the fundamental atmospheric parameters of HD~49434
from the Str\"omgren photometry given in Table~\ref{stromgren_tab}.
We have used the {\sc templogg} code (Rogers~1995, 
see also~Kupka \& Bruntt~2001) and we
find the atmospheric parameters of HD~49434 to be
$T_{\rm eff} = 7\,300\pm200$~K, $\log g = 4.14\pm0.20$, 
and [Fe/H]$ = -0.01\pm0.20$.
The quoted errors are approximate and are dominated by 
the uncertainty from the photometric calibrations.

We determined the microturbulence parameter to be 
$\xi_t = 2.1\pm0.5$ \kms\ (cf.~discussion in Section~\ref{abund_error}). 
We used macroturbulence $v_{\rm macro} = 5.0$ \kms\ and 
$v \sin i = 84\pm4$ \kms. The choice of $v_{\rm macro}$ is
not important here, since the width of the lines is dominated by the
rotational broadening.

Several lines fitted by \vwa\ are shown in Figure~\ref{lines_fig}.
We have also shown the relative difference between the synthetic and observed
spectrum. Neighbouring lines with a line depth of at least 
25\% of the fitted line are also shown. 
The length of the tick marks show the relative line depth
of each neighbouring line compared to the fitted line.

From Figure~\ref{lines_fig} it can be seen that several lines 
that were used for the abundance analysis are blended.
Due to the high \vsini\ of HD~49434 we simply had to 
use the least blended lines. 
Our strategy was to determine the Fe abundance accurately from
non-blended lines. Consequently, we held the abundance of Fe fixed, 
and allowed \vwa\ to select lines of other elements that were mildly 
blended by Fe lines, eg.~the Ca lines at 5594.5 and 6102.7~\AA.
More extreme examples are the C, Ni, and Sc lines at 5052.2, 5614.8, and
5526.8~\AA~where the lines are heavily blended and the abundance 
determination is unreliable, i.e.~the $S_i$ are 0.49, 0.24, and 0.22. 
If we again assume that the equivalent 
width is known to 5\% the corresponding error is 0.1 dex for the C lines
and 0.2 dex for the Ni and Sc lines. In addition, it is difficult to 
establish the location of the continuum level because the lines are so broad.
Realistic error estimates lie in the range 0.2--0.4 dex for these three lines.
Note that we did not find enough lines to give a reliable estimate for
the abundance Ni and for C we found only one line that was usable.

The result of the abundance analysis is given in Table~\ref{abundances_tab} 
and for comparison we also give the solar abundances 
from Grevesse \& Sauval \cite{sunabund}. The abundances and errors
are calculated using weights that are estimated from the
errors on the individual lines (cf.~Section~\ref{sensitivity}).
The error estimates that we give in Table~\ref{abundances_tab} 
do not include the systematic error due to the uncertainty of the 
atmospheric model parameters which will be discussed in 
Section~\ref{abund_error}.

Generally we find that the abundances of most elements lie around 
the Solar value within the error bars. 
Only for Fe and Ca do we find enough lines (16 and 7) 
to make a truly reliable estimate of the abundances.


\begin{table} \centering \footnotesize
  \caption{Str\"omgren photometric indices taken from 
Hauck \& Mermilliod (1998) for
the four stars used for the HD~49434 observations from 
Sierra Nevada\label{stromgren_tab}}
  
  \begin{tabular}{c|ccccc}
 \hline 
  Star   & $V$  &  $b-y$    & $m_1$    & $c_1$  & ${\rm H}_\beta$ \\
 \hline
HD~48922 & 6.77 &   -0.015  &   0.142  & 0.940 & -     \\
HD~49933 & 5.78 &    0.270  &   0.127  & 0.460 & 2.662 \\
HD~50747 & 5.45 &    0.095  &   0.172 & 1.131  & 2.833 \\
  \hline
HD~49434 & 5.74 &    0.178  &   0.178  & 0.717 & 2.755 \\ 

  \end{tabular}
\end{table}


The Fe abundance is $\log N_{\rm Fe}/N_{\rm tot} = -4.67\pm0.02$ from
16 Fe\iet\ lines (we quote the internal weighted error; the internal RMS
scatter is 0.18~dex).
If we assume that the photospheric hydrogen and helium abundance is 
the same in HD~49434 and the Sun this abundance 
corresponds to [Fe/H] $= -0.13\pm0.14$ (including systematic errors). 
This agrees roughly with the result from the Str\"omgren $m_1$
index which yields [Fe/H] $= -0.01\pm0.20$.

An interesting result is the ratio [Ca/Fe] $= +0.30\pm0.21$
which is a quite high value for a star with Solar-like abundances.

We note that we have adjusted the $\log gf$ for the Ca\ii\ line at 
5857.451 \AA~line by comparing with the solar spectrum, 
i.e.\ from $\log gf = 0.257$ to $0.550$. 
The $\log gf$ values for the Si lines are determined solely from 
theoretical calculations. Therefore we have computed the
synthetic spectrum for the Si-lines we have used -- using the
solar abundance -- and comparing this with the observed solar spectrum. 
We find that the $\log gf$ values seem to be right for the lines
we have used.

Lastennet et al.\ \cite{lastennet} have determined fundamental
atmospheric parameters of HD~49434 by comparison of the observed and 
synthetic photometric colours. The star is among the nine 
potential COROT targets stars they have studied. 
From the combination $(B-V)$, $(U-B)$, $(b-y)$ they find
$\log g = 4.0\pm0.4$, $T_{\rm eff} = 7\,240\pm100$~K and [Fe/H]$=-0.1\pm0.2$
(from Table 2 and Figure 7 and 8 in Lastennet et al.\ 2001).
Thus, the metallicity they find is in agreement with our result from
spectroscopy and all the fundamental parameters are in agreement
with the estimates based on Str\"omgren photometry.

We finally note that HD~49434 is included in the
$\Delta a$ photometry catalogue of bright B and A stars by 
Vogt et al.\ (1998) but they find no evidence for the 
star being chemically peculiar.

\subsubsection{Accuracy of the derived abundances\label{abund_error}}

The errors quoted in Table~\ref{abundances_tab} for the abundances are the 
weighted internal errors (cf.~Section~\ref{abund_error}). Here we will
discuss the error contribution from the uncertainty of the atmospheric 
model parameters. The atmospheric model we used for the final abundance
result have the parameters $T_{\rm eff} = 7\,300$~K and $\log g = 4.1$.
To investigate the effect of the choice of atmospheric model we calculated 
Fe abundances for models with $T_{\rm eff} =7\,100, 7\,300, 7\,500$~K 
and $\log g = 4.1$ and one model 
with $T_{\rm eff} = 7\,300$ and $\log g = 4.3$.
We find that the effect of increasing $\log g$ by $0.2$~dex decrease the
Fe abundance by about $0.04\pm0.05$~dex (sd.~of mean) which 
is not significant.
On the other hand, if we increase the temperature of the
atmospheric model by $200$~K we find an increase 
of Fe by $0.10\pm0.05$~dex, i.e.~a significant effect. 

Another input parameter for the calculation of the synthetic spectrum 
is the microturbulence. 
The value used in the calculation of the Kurucz atmosphere model
does not affect the derived abundances significantly,
but when calculating the synthetic spectrum the effect is indeed
significant. When changing 
the microturbulence the change in equivalent width 
depends on the strength of the line. By using this fact, one
can adjust the microturbulence until the abundance of individual
lines do not correlate with equivalent width. We could not do this
very accurately due to the high rotational velocity of HD~49434.
Thus, we estimate the microturbulence to be $\xi_{\rm t} = 2.1\pm0.5$ \kms.

We find that when increasing $\xi_{\rm t}$ by 0.5 \kms\ the abundance
of Fe decreases by $0.10\pm0.04$~dex, i.e.~a significant
contribution to the uncertainty of the abundance.

We conclude that the contribution to the error of 
abundances due to uncertain 
model atmosphere parameters, i.e.~$T_{\rm eff}$ and
microturbulence, is of the order 0.14 dex. For a realistic
estimate of the abundance of elements in HD~49434 this 
contribution must be added to the internal errors (sd.~of mean)
given in Table~\ref{abundances_tab}. For example for the metallicity
of HD~49434 we get [Fe/H] $= -0.13\pm0.14$, i.e.~the uncertainty of
the model parameters is the main error source.


\begin{figure}\centering
  \begin{center}
    \caption{
Power spectra corresponding to the variable star HD~49434 minus the
comparison star HD~48922, in the upper panel, and that corresponding to the
comparison stars HD~49933 minus HD~48922 in the bottom panel. 
The scales are milli-magnitudes and cycles/day\label{strom2_fig}}
 \includegraphics[angle=270,width=7.4cm]{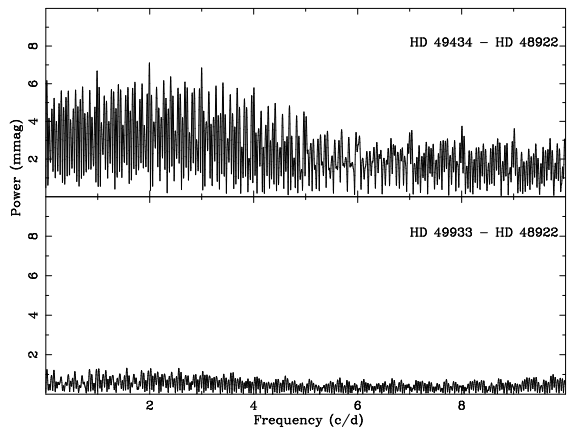} 
  \end{center}
\end{figure}

\section{Str\"omgren photometry time series \label{photometry}}

We have observed HD~49434 with a Str\"omgren photometer on 
the 0.9m~telescope at Sierra Nevada near Granada, Spain. 
We used a $45"$ diaphragm to observe HD~49434 and the comparison stars
HD~48922=C1, HD~49933=C2, and HD~50747=C3. 
The star HD~49933 is a V = 5.78 double system with a
faint companion of magnitude V = 11.
This last star was included within the diaphragm because of its
proximity but had no influence on the final
differential photometry. The Str\"omgren parameters for the observed
stars are given in Table~\ref{stromgren_tab}.

\begin{figure}
  \begin{center}
    \caption{
Location of HD~49434 (star symbol) in the HR diagram together
with the known $\delta$ Scuti (circles, Rodr\'\i{}guez \& Breger 2001) 
and bona fide $\gamma$~Dor (triangles, Handler \& Kaye 2001) pulsators. 
The edges of the $\gamma$~Dor region around $(b-y)_0=0.2$ are 
from Handler (1999). 
The absolute magnitude of HD~49434 is $M_V = 2.73\pm0.07$ and 
is derived from its Hipparcos parallax (ESA 1997)\label{hr_diag_hd49434}}
 \includegraphics[height=8.4cm,width=5.0cm,angle=270]{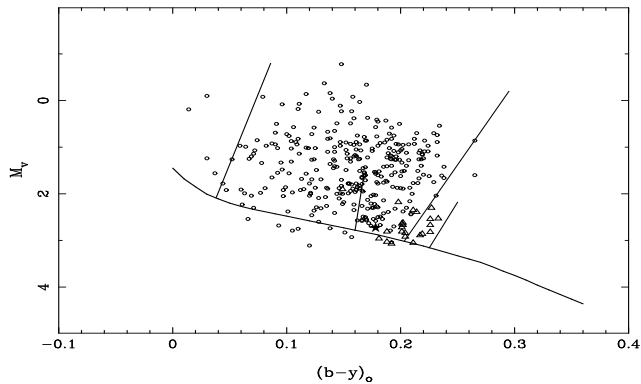}
  \end{center}
\end{figure}

Continuous photometric time series of comparison stars
HD~48922, HD~49933 and HD~49434 were made in order to remove 
the sky transparency fluctuations to be able to see 
the variations of the target star.
Magnitude differences between HD~49933 and HD~48922 are constant 
within 3.7 mmag (rms) and clear intrinsic variations can be 
seen for HD~49434. The light curve is shown in Figure~\ref{strom1_fig}.

The time series consist of data from nine nights spread unevenly 
over one month starting on July 2001.
Unfortunately the weather conditions were not good during our 
first night of observation and HD~49434 was observed again a week later. 
Observations were possible only three weeks later where the star 
was included in a program where other targets were also measured. 
Thus the dedicated observing time was limited to only a few
measurements but the data could be used to confirm the long-term 
variations that were seen in the first days.

In total the time series consists of 295 data points
in the four Str\"omgren bands ($uvby$). In Figure~\ref{strom1_fig} we show
the time series of HD~49434 - HD~48922 (target - comparison star 1) and 
HD~49933 - HD~48922 (comparison star 2 - comparison star 1).
The magnitude differences are plotted for the Str\"omgren $b$ band.

\begin{figure*}[t]
\includegraphics[width=8cm,height=8cm]{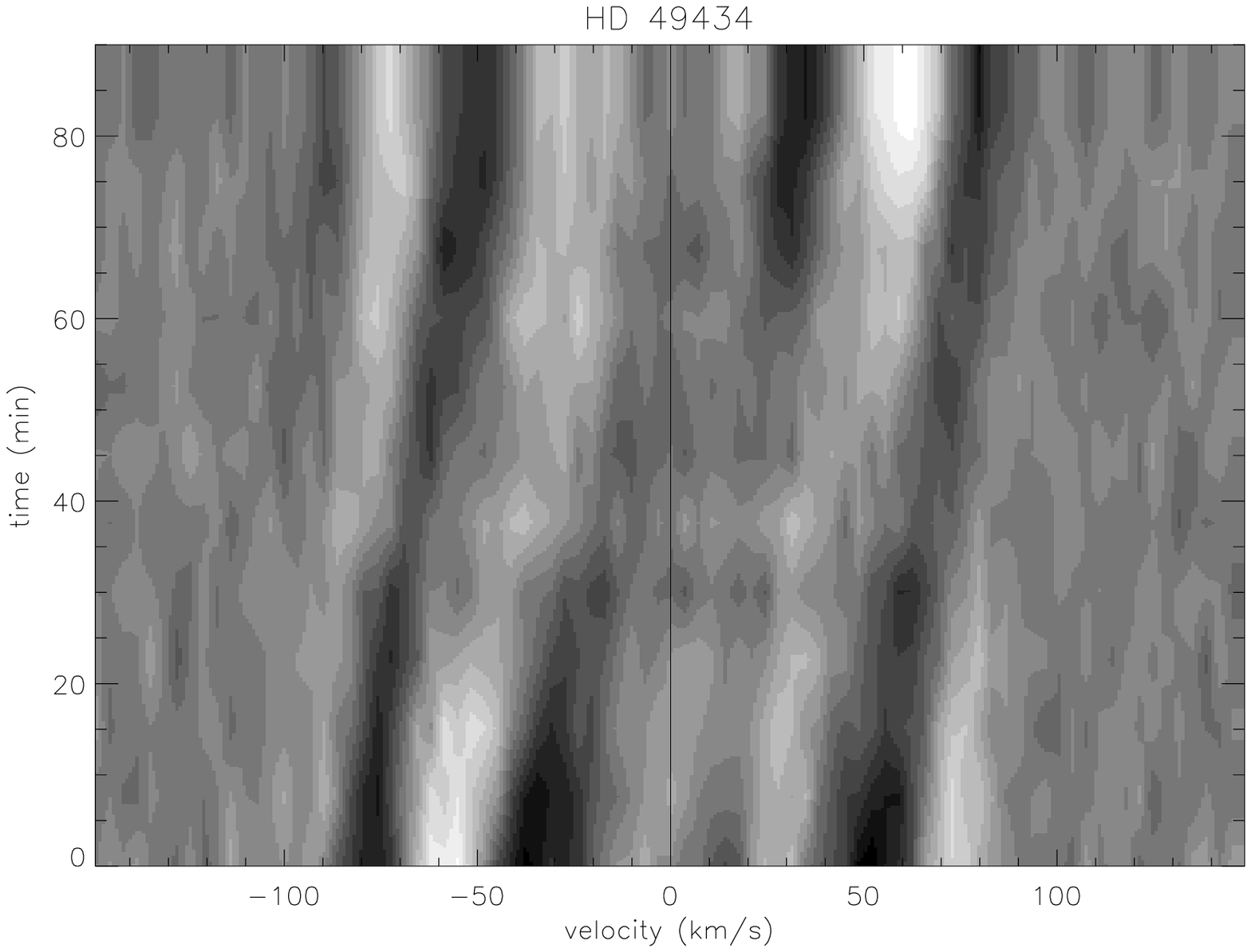}
\hskip 0.5cm
\includegraphics[width=8cm,height=8cm]{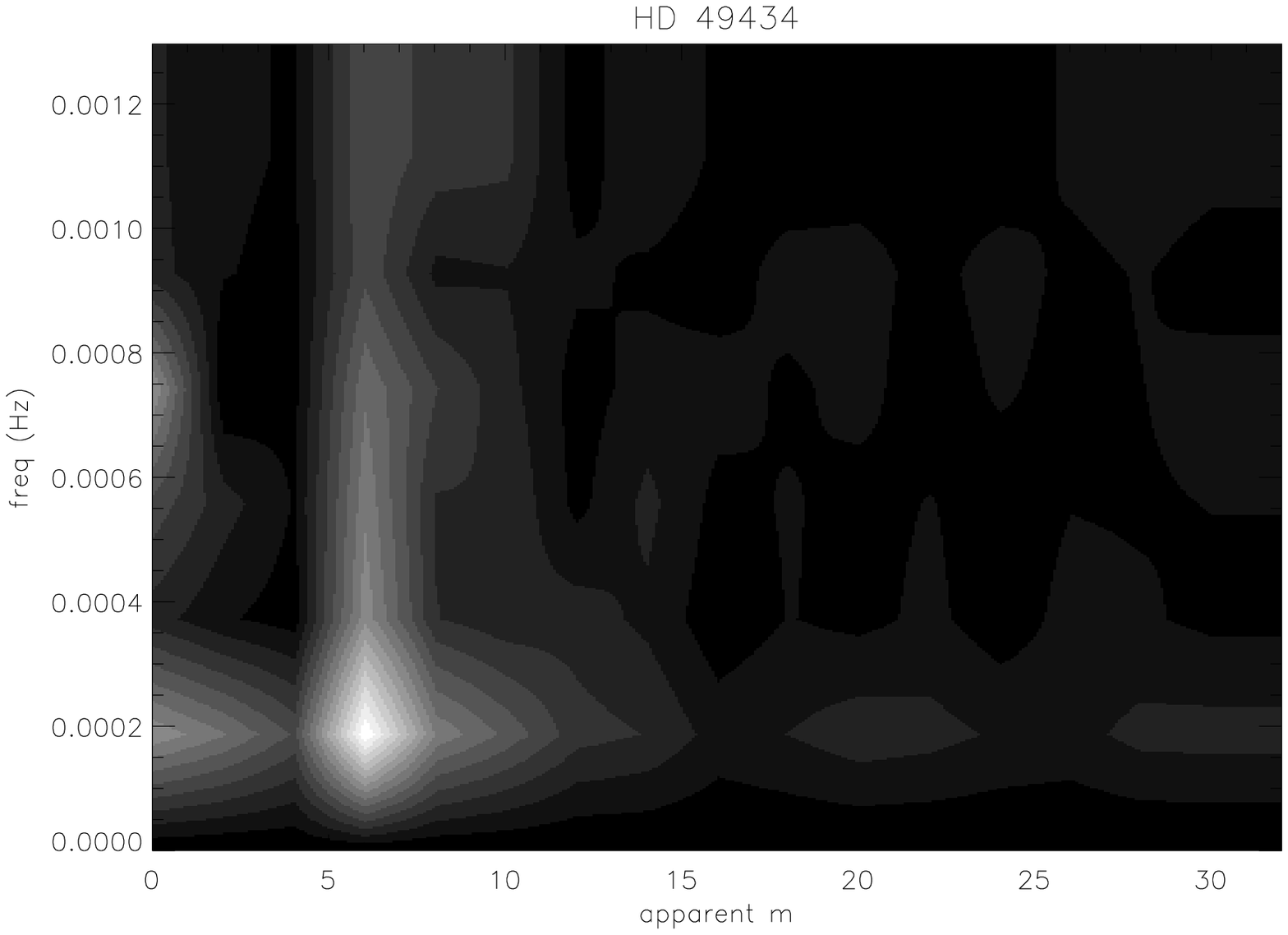}
\caption[]{The residual of the mean profile for the January 2000 series. 
{\it Left:} in velocity -- time space; the maximum amplitude is of the order 
of $3 \times 10^{-3}$; {\it Right:} Spectral power density of the same data,
in apparent order -- frequency space, after the 2D Fourier transform}
\label{fourier1}
\end{figure*}

\begin{figure*}[t]

\includegraphics[width=8cm,height=8cm]{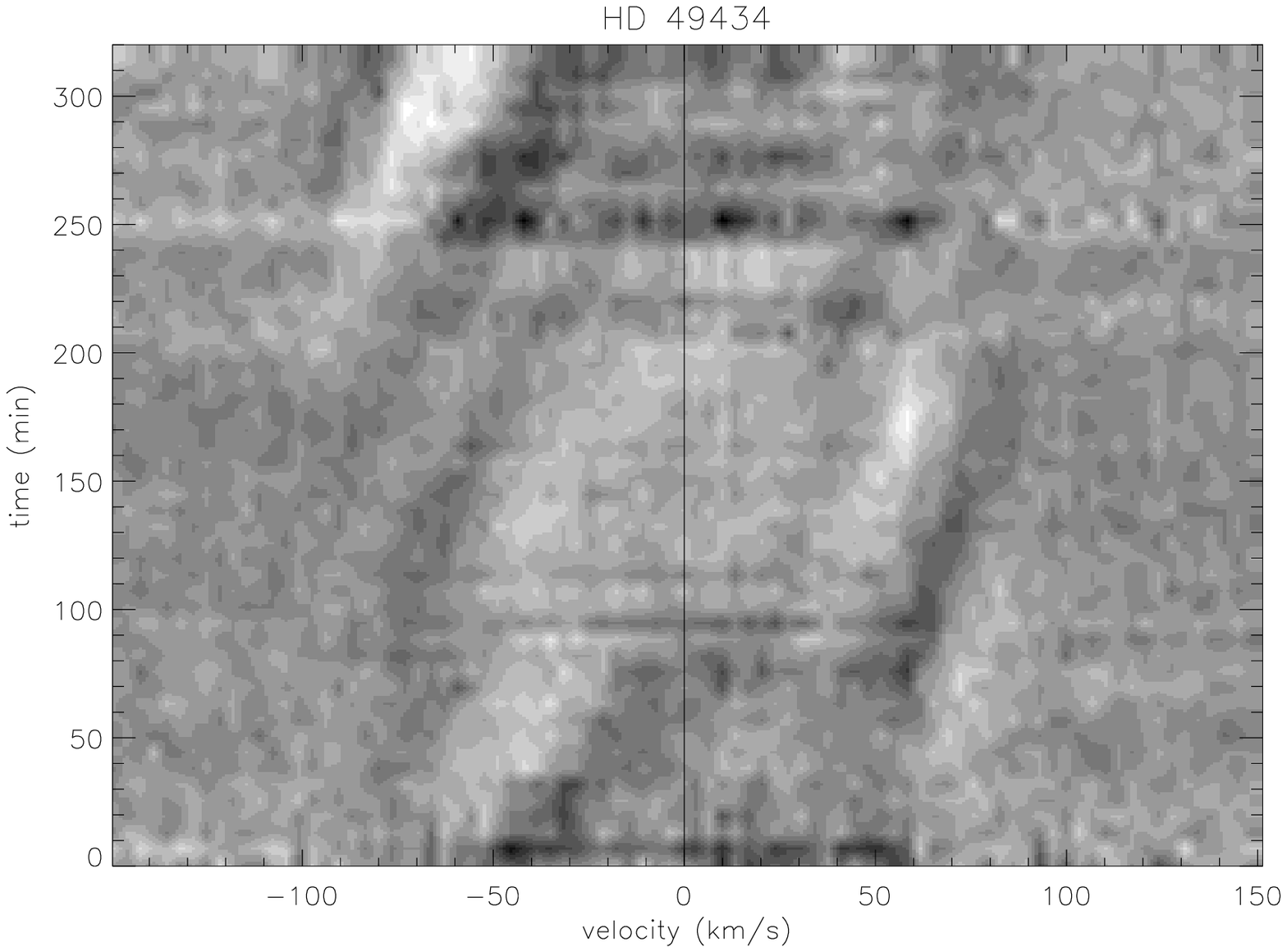}
\hskip 0.5cm
\includegraphics[width=8cm,height=8cm]{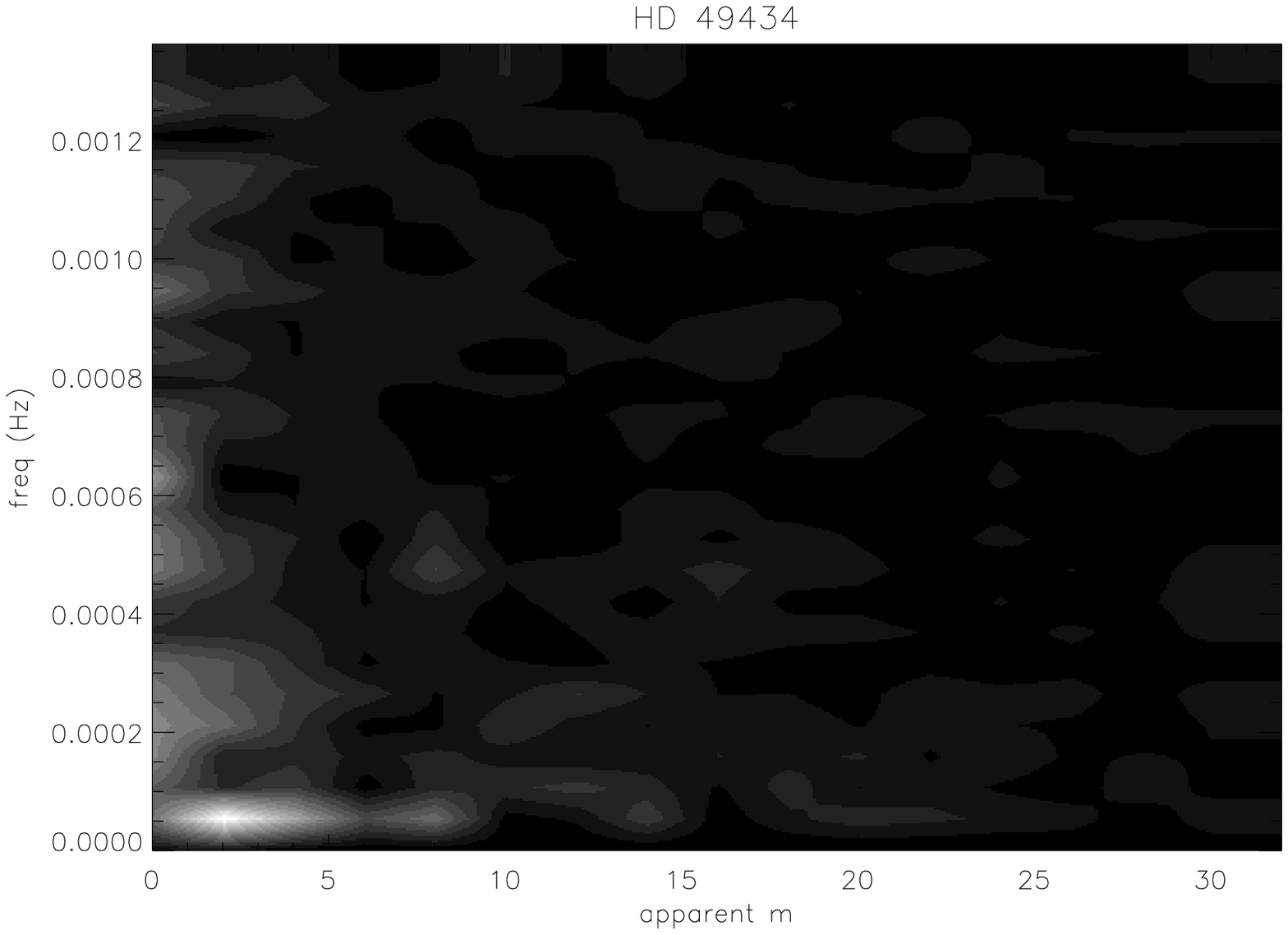}

\caption[]{The same as Figure~\ref{fourier1} but for the December 2000 series; the maximum
amplitude of the residuals is of the order of  $4 \times 10^{-3}$ 
\label{fourier2}}
\end{figure*}

From Figure~\ref{strom1_fig}  it is clear that the 
intrinsic photometric variations are
associated with HD~49434. At first sight the star does not 
seem to have periodic variations. A power spectrum was 
calculated in order to search for
possible frequency components which is shown in Figure~\ref{strom2_fig}.
It can be seen that with the present data, no clear discrete 
frequency components are present. Instead the low frequency region 
has a significant contribution for HD~49434 which is not
seen for the comparison stars.

This behaviour is typical among the recently discovered
$\gamma$ Dor stars (Zerbi et al.\ 1999) which are g-mode pulsators.
We believe that HD~49434 is a variable star of this type:
the star is classified as a main sequence F2-type star 
from Str\"omgren photometry and its position in the
Hertzsprung-Russell diagram in Figure \ref{hr_diag_hd49434} confirm 
our hypothesis.
Because we have not detected any clear periodicity it is not possible to test
the phase differences among the different photometric colours which is a
clear signature of the non-radial pulsating nature of the $\gamma$ Dor
stars -- see Garrido~\cite{garrido2000} for details.


\section{Spectroscopic monitoring of HD~49434\label{lsd}}

In the following we will discuss the analysis of the 
time series of spectra of HD~49434 to look for any systematic 
variation of line profiles.

\subsection{Data reduction\label{lsd-datareduction}}

All data were reduced on-site, using the automatic INTER-TACOS procedure 
(Baranne et al.\ 1996), resulting in wavelength-calibrated, 
flat-field corrected spectra. 
Each spectrum is segmented into 67 segments corresponding to the 
67 orders of the spectrograph present on the detector. We then performed a 
large scale normalisation of the overall spectrum, so that the resulting 
spectra are normalised to unity all across the covered wavelength domain,
even though some local problems still remain with this normalisation on the
edges of the orders.

The ELODIE spectrograph provides a wide enough spectral coverage to give 
access to a large number of photospheric lines. In order to take advantage
of this, and to fully exploit the information contained in these numerous
photospheric lines, we used the Least-Square Deconvolution (LSD) technique 
(Donati et al.\ 1997) to analyse the variations of a ``mean'' photospheric 
line. In this method, a line pattern function is constructed, containing all 
the lines supposedly present in the spectrum as Dirac functions, with heights 
set to the central line depths as calculated by Kurucz's (1979) ``SYNTHE'' 
program. The observed spectrum is then deconvolved with this line pattern 
function, yielding a ``mean'' photospheric line profile. With this technique, 
line blends are automatically taken into account when all the lines present 
in the spectrum are considered. Note that the depth of the resulting ``mean'' 
line has no physical meaning, but that time variations of this depth, as 
well as line profiles can be accurately analysed with this technique. 
We used a Kurucz model for $T_{\rm{eff}} = 7\,000$~K and log~$g = 4$, 
sufficiently close to the values derived in this paper, for constructing the 
line pattern function. 

A total of 1700 lines were finally used in this analysis. In the following of
this work, we shall call ``mean'' photospheric line, the line constructed with
the LSD technique, while the adjective ``average'' will be reserved for time 
averages of line profiles.

All mean profiles in each series were normalised to the same equivalent 
width. This procedure eliminates the effects of the large scale variations
of the mean line profile, due for instance to normalisation problems that may
vary from spectrum to spectrum. The time average of the mean profiles was
then constructed for each series, and subtracted from the mean line profiles
of the series.

We then performed a two-dimensional Fourier transform analysis, following 
the method described in Kennelly et al.\ (1996). This method is well suited
for the analysis of the spectroscopic signature of non-radial pulsations.
Each line profile was interpolated on a grid representing stellar longitudes, 
transforming velocities across the line profile into longitudes on the stellar
equator using the relation $v-v_r\; =\; v \sin i\; \sin \phi$, where $v$ 
is the velocity in the profile, 
$v_r$ is the radial velocity of the star, 
measured to be $v_{\rm rad} = -12.3\pm1.0$ \kms, 
$\phi$ is the corresponding stellar longitude, 
and $v \sin i$ is the projected rotation velocity of the star, 
estimated to be $84\pm4$ \kms.

For the time series from January 2000, the spectra were recorded at constant time 
intervals during the whole moni\-toring, while a few short gaps exist 
in the December 2000 series. For this latter series, prior to the 2D Fourier
transform, we had to build up a grid of constant time intervals, filling the
gaps during which no spectrum was obtained with null data.

After these steps, the data were submitted to a 2D Fourier transform, 
resulting in a representation of the data in the (frequency -- apparent $m$) 
space. The apparent $m$ is related to the structure of the modes
present at the stellar surface, without being identical to the usual azimuthal 
order $m$, since the apparent $m$ depends for instance on the inclination 
angle $i$ of the rotation axis of the star with respect to the line of sight. 
The apparent $m$ is nevertheless indicative of the surface structure of 
the modes detected.

\subsection{Detected variability of spectral lines}

Figure~\ref{fourier1} (and Figure~\ref{fourier2}) presents the results 
of this analysis for the series from January 2000 (and December 2000).
At both epochs we detect the presence of line profile perturbations, crossing
the profile from blue to red. Such perturbations are most probably the 
signature of non-radial pulsation modes at the stellar surface. 
The Figures~\ref{fourier1} and \ref{fourier2}  
suggest the presence of one or several high-degree modes. 
These modes are particularly obvious in the January 2000 series, owing to
the high signal-to-noise ratio of the spectra, while they are less 
conspicuous, although present without doubt, in the lower quality December
2000 data. The amplitudes of the residuals are of the order of 
2--3$\times 10^{-3}$ at both epochs. They can be easily detected thanks to the
very high signal-to-noise ratio (S/N) of the LSD mean line profiles combining 
hundreds of spectral lines, which reach S/N=3\,000 in the December 2000 
series and S/N=6\,000 in the January 2000 series.

These modes seem to be long-period modes.
This is particularly clear in Figure~\ref{fourier2} (right), where the
frequency of the detected modes are at the resolution limit of our dataset,
i.e.~0.1~mHz, indicating a period longer than 160 minutes, that is to say 
longer than half the time span of the monitoring. These results therefore
confirm those of the photometric observations, and suggest that 
HD~49434 is probably a variable of the $\gamma$ Doradus type.

The modes we detect in the January 2000 series are high-degree modes, with
an apparent order of 6. In the December 2000 series, our lower quality 
data still indicates non-radial modes, but with more power on the apparent
order 2, some signal still appearing at apparent order 8.

\section{Conclusions and future prospects}

We have presented the detailed abundance analysis of HD~49434.
The analysis could not be made by simple equivalent width
measurements since most lines are blended, because
the star has high projected rotational velocity. 
We have described the \vwa\ software which 
selects the least blended lines and derive abundances 
semi-automatically. 

We find that the metallicity of HD~49434 is somewhat below the 
Solar value, i.e.~[Fe/H]$ = -0.13\pm0.14$. The quoted error includes 
the estimated error due to the uncertainty of the atmosphere 
parameters, especially $T_{\rm eff}$ and microturbulence.

The metallicity we find agrees with the results from Str\"omgren 
photometry and the work by Lastennet et al.\ \cite{lastennet}
who have compared observed and synthetic photometric colours.
When taking into account the uncertainty of the basic atmosphere 
parameters the accuracy of \vwa\ is only slightly more accurate 
than the photometric methods. The advantage of analysing the spectra
with \vwa\ is that we obtain estimates for several elements. 

We have already used \vwa\ for the analysis of 
several other stars which will be published soon 
(Bikmaev et al.\ 2002, Rasmussen et al.\ 2002).
In the paper by Bikmaev et al.\ we have analysed two A-type stars, 
which have low $v \sin i$. Thus we have made a detailed comparison 
of the result obtained with \vwa\ and with the ``classical'' method 
based on equivalent width measurements of non-blended lines -- the
agreement is excellent for nearly all lines (abundances $<0.03$~dex).

With the confirmation that \vwa\ can be used for reliable 
semi-automatic abundance analysis we will proceed to use the 
software for analysis of the proposed main targets for the 
asteroseismology satellite missions COROT and MONS/R\o mer: 
the spectra have already been obtained.

For our future use of \vwa\ we expect that
for stars with low \vsini\ ($<$ 40--50 \kms) it will be possible to 
constrain the atmospheric parameters 
(eg.\ microturbulence and $T_{\rm eff}$) by adjusting these 
so the abundances found from lines of the same element 
do not correlate with the line parameters 
(eg.\ equivalent width and excitation potential).
For stars with higher \vsini\ ($> 100$ \kms) the selection
of atomic lines for abundance analysis becomes exceedingly difficult. 
For such stars the metallicity and the basic parameters 
must be determined mainly from photometric colours, but
it is still important to obtain spectra to be able to put constraints
on $T_{\rm eff}$ (from the hydrogen lines) and \vsini. 
For example, stars with too high \vsini\ may not be 
suitable for seismology since this will make the analysis 
much more complex.

In the present paper we have also presented the photometric
light curve of HD~49434 based on observations during part of nine 
nights over a period of one month. We find that the star is indeed
variable with excess power around 1-5 cycles per day, 
but no periodic signal could be found from the limited data set.
We have also analysed the spectroscopic line profiles which show
evidence for long-period variability at low amplitude. 
This behaviour is characteristic for the $\gamma$ Dor 
variables, and we propose that HD~49434 is indeed a variable of this 
type. A longer time series is needed to be able to detect any 
periodic pulsation signal. 

The star has been chosen as a main target for COROT which 
means that the star will be observed for around 150 days. 
We will then be able to probe the interior of this star in detail
by using seismic techniques. To do this one needs to
constrain the fundamental parameters of the star which indeed has 
been aim of the work presented here.


\begin{acknowledgements}

We wish to thank Tanya Ryabchikova and Christian St\"utz 
for advice on the development of the \vwa\ program and to 
Werner W.~Weiss at the Astronomy Institute of the University of 
Vienna for letting us use the \vald\ database.
We also wish to thank Nikolai Piskunov for letting us use his 
software for the calculation of synthetic spectra ({\sc synth}).
Thanks to R.~L.~Kurucz for making his model atmosphere code and
data available.

\end{acknowledgements}


\end{document}